\documentclass[12pt,preprint]{aastex}
\def\degpoint{\ifmmode ^{\rm{o}}\!. \else $^{\rm{o}}\!.$\fi}

\newcommand{\ms}{\mbox{m\,s$^{-1}$}}
\newcommand{\kms}{\mbox{km \ s$^{-1}$}}
\newcommand{\Msun}{\mbox{M$_{\odot}$}}
\newcommand{\Rsun}{\mbox{R$_{\odot}$}}
\newcommand{\Mjup}{\mbox{M$_{\rm Jup}$}}

\newcommand{\Lsun}{\mbox{L$_{\odot}$}}
\newcommand{\Mearth}{\mbox{M$_{\oplus}$}}

\newcommand{\ltsimeq}{\raisebox{-0.6ex}{$\,\stackrel
         {\raisebox{-.2ex}{$\textstyle <$}}{\sim}\,$}}
\newcommand{\gtsimeq}{\raisebox{-0.6ex}{$\,\stackrel
         {\raisebox{-.2ex}{$\textstyle >$}}{\sim}\,$}}

\begin{document}

\title{The Pan-Pacific Planet Search. I. A Giant Planet Orbiting 7 CMa }

\author{Robert A.~Wittenmyer\altaffilmark{1}, Michael 
Endl\altaffilmark{2}, Liang Wang\altaffilmark{3}, John Asher 
Johnson\altaffilmark{4}, C.G.~Tinney\altaffilmark{1}, 
S.J.~O'Toole\altaffilmark{5} }
\altaffiltext{1}{Department of Astrophysics, School of Physics, 
University of NSW, 2052, Australia}
\altaffiltext{2}{McDonald Observatory, University of Texas at Austin, 1 
University Station C1400, Austin, TX 78712, USA }
\altaffiltext{3}{Key Laboratory of Optical Astronomy, National 
Astronomical Observatories, Chinese Academy of Sciences, A20 Datun Road, 
Chaoyang District, Beijing 100012, China}
\altaffiltext{4}{Department of Astrophysics, California Institute of 
Technology, MC 249-17, Pasadena, CA 91125, USA}
\altaffiltext{5}{Australian Astronomical Observatory, PO Box 296, 
Epping, 1710, Australia}

\email{
rob@phys.unsw.edu.au}

\shorttitle{Planet Around 7 CMa }
\shortauthors{Wittenmyer et al.}

\begin{abstract}

\noindent We introduce the Pan-Pacific Planet Search, a survey of 170 
metal-rich Southern hemisphere subgiants using the 3.9m Anglo-Australian 
Telescope.  We report the first discovery from this program, a giant 
planet orbiting 7 CMa (HD~47205) with a period of 763$\pm$17 days, 
eccentricity $e=0.14\pm$0.06, and m~sin~$i$=2.6$\pm$0.6\,\Mjup.  The 
host star is a K giant with a mass of 1.5$\pm$0.3\,\Msun and metallicity 
[Fe/H]=0.21$\pm$0.10.  The mass and period of 7~CMa~b are typical of 
planets which have been found to orbit intermediate-mass stars 
($M_{*}>1.3$\Msun).  \textit{Hipparcos} photometry shows this star to be 
stable to 0.0004 mag on the radial-velocity period, giving confidence 
that this signal can be attributed to reflex motion caused by an 
orbiting planet.

\end{abstract}


\section{Introduction}

Nearly 20 years of concerted radial-velocity monitoring of solar-type 
main-sequence stars has unveiled a fascinating diversity of planets and 
planetary system configurations.  From the many hundreds of planets now 
characterised, the observational evidence is mounting for several 
interesting relationships between the properties of planets and their 
host stars.  Among these are: (1) giant planet occurrence is positively 
correlated with stellar metallicity \citep{fv05} and mass 
\citep{johnson10a, endl06}, (2) short-period ``Super-Earths'' with 
m~sin~$i<$10\,\Mearth\ are about an order of magnitude more common than 
close-in giant planets \citep{howard10, wittenmyer11}, and (3) planet 
mass is positively correlated with host star mass \citep{bowler10}.  

Most of the stars which have been targeted by radial-velocity surveys 
have masses which fall in the range 0.7-1.3 \Msun\ \citep{johnson07, 
vf05}.  This is a consequence of the technical requirements of Doppler 
exoplanetary detection, which demand that stars be cool enough to 
present an abundance of spectral lines, and rotate slowly enough that 
their absorption lines are not significantly broadened by rotation.  
Stars of lower mass (e.g.~M dwarfs) are intrinsically faint in the 
optical, making the acquisition of high signal-to-noise spectra 
extremely expensive in telescope time \citep{endl06}.  Main sequence 
stars of higher mass have few usable absorption lines (due to their high 
temperatures), and also tend to be fast rotators ($v\sin~i>$ 50 \kms; 
Galland et al.~2005) due to their youth.  In addition, the shorter 
main-sequence lifetimes of higher-mass stars means that they will 
preferentially be observed at younger ages.  Stars earlier than about F7 
also have much shallower convection zones, and so do not experience the 
magnetic braking which slows the rotation of later-type (lower-mass) 
stars.  As a result, only the most massive planets can be detected 
orbiting A and F dwarfs.  It is only recently that a significant number 
of planetary systems have been discovered orbiting intermediate-mass 
stars ($M_{*}>1.3$\Msun).  These stars have proven to be a fertile 
hunting ground for interesting planetary systems, such as the 4:3 
mean-motion resonant planets orbiting HD~200964 \citep{johnson11}.  Now, 
some headway is beginning to be made in addressing the crucial question 
of how planet formation depends on stellar mass \citep{bowler10, 
johnson10a, sato10}.  A number of surveys are seeking to expand our 
knowledge of planetary systems orbiting stars more massive than the Sun, 
e.g.~\citet{setiawan03, hatzes05, sato05, johnson06b, doellinger07, 
n09}.  These surveys are exploiting the advantage wrought by stellar 
evolution: as stars evolve off the main sequence into subgiants and 
giants, their atmospheres expand and cool, making precision Doppler 
velocity measurements possible due to an abundance of narrow spectral 
lines.

The well-known planet-metallicity correlation \citep{gonzalez99, fv05}, 
whereby main-sequence stars with higher metal content are more likely to 
host planets, has come under some scrutiny.  Analysis of the 
metallicities of planet-hosting giant stars by \citet{schuler05} showed 
that the giant star planet hosts were significantly more metal-poor than 
their main-sequence counterparts.  \citet{pasquini07} have also argued 
that the planet-metallicity correlation does not apply for evolved 
stars.  They propose that this is evidence for a ``pollution'' scenario, 
in which main-sequence stars hosting planets appear metal-rich because 
they have accreted material from the protoplanetary disk 
\citep{murray02}.  When a star evolves off the main sequence, the 
convective zone increases in size by about a factor of 35 
\citep{pasquini07}.  If the high metallicities observed in planet hosts 
are due to pollution, this expansion of the convective zone will 
significantly dilute the extent of that pollution, and the subgiant's 
photosphere would return to its ``birth'' metallicity.  Hence, one would 
\textit{not} expect a significant correlation between metallicity and 
planet frequency for subgiants.

However, the importance of planet pollution was downplayed even by the 
authors who proposed it, as they felt it should play only a minor role 
in shaping the planet-metallicity correlation seen in dwarf stars.  
Further, Valenti \& Fischer~(2008), Johnson et al.~2010a, Takeda et 
al.~(2007) and others found no evidence of a decreasing 
planet-metallicity correlation among F dwarfs, subgiants or K giants.  
The sample of K giants studied by \citet{pasquini07} had a limited 
metallicity range, with [Fe/H]$< +0.2$.  Examination of the form of the 
planet-metallicity correlation of \citet{fv05} and \citet{johnson10a} 
shows that, for small numbers of stars, the correlation over this 
metallicity range would look approximately flat.

One way to test this is to search for planets around a sample of evolved 
stars that are unambiguously metal rich.  \citet{sandage03} pointed out 
that stars on the red-edge of the subgiant branch represent such a 
population.  Here, we introduce a new Southern Hemisphere survey, the 
``Pan-Pacific Planet Search,'' which uses the 3.9-metre Anglo-Australian 
Telescope (AAT) to search for planets among these evolved, metal-rich 
stars to test for evidence of planet pollution.

In this paper, we present the first result from this new survey: the 
detection of a 2.6 \Mjup\ planet orbiting the nearby evolved star 
HD~47205.  In Section~2, we introduce the Pan-Pacific Planet Search and 
give a complete target list for the survey.  Section~3 describes the 
observational data and gives the stellar parameters for 7~CMa.  In 
Section~4, we detail the orbit-fitting process and present the planetary 
parameters.  Finally, in Section~5 we discuss the further implications 
of this discovery.

\section{The Pan-Pacific Planet Search}

\subsection{Survey Strategy and Target Selection}

The Pan-Pacific Planet Search (PPPS) originated as a Southern hemisphere 
extension of the established Lick \& Keck Observatory survey for planets 
orbiting Northern ``retired A stars'' \citep{johnson06b, johnsonetal07, 
johnson10a}.  This program is using the 3.9m Anglo-Australian Telescope 
(AAT) to observe a metal-rich sample of Southern Hemisphere subgiants.  
We have selected 170 Southern stars with the following criteria: $1.0 < 
(B-V) < 1.2$, $1.8 < M_V < 3.0$, and $V<8.0$.  By requiring $(B-V)>1$, 
we extend the red limit of the \citet{johnson06b} survey to the colours 
that stellar models indicate will be dominated by metal-rich subgiants 
\citep{girardi02}.  This aims to deliver improved planetary detection 
statistics at [Fe/H]$>$0.0.  In light of the observed positive 
correlation between stellar metallicity and planet occurrence, this 
should also deliver a roughly equivalent number of planetary detections 
to that obtained at Lick and Keck, though for metal-rich hosts.  At the 
same time, by requiring $M_V>1.8$, we exclude giant-branch stars, as 
these have significant intrinsic velocity noise (``jitter'') due to 
random convective motion and pulsations \citep{saar98, wright05} -- 
typically about 20 \ms\ \citep{hekker06}.  Our target list includes 
about 30 stars from the Lick survey; this overlap will serve as a check 
on the systematics between the two telescopes.  Together, the three 
telescopes are observing more than 600 stars.  The complete PPPS target 
list is given in Table~\ref{targetlist}.

Observing time is scheduled such that each target should receive 4--6 
observations per year.  This strategy would appear to reduce the 
probability of detecting shorter-period planets ($P\ltsimeq$50 days), 
which require more densely-sampled observations in continuous blocks of 
time \citep{16417paper, monster, vogt10}.  However, we note that the 
same scheduling has been used in the Lick and Keck survey, which has 
detected the $P=6.5$ day planet orbiting HD~102956 \citep{johnson10c}.  
By employing this strategy, we are able to target more stars with a 
fixed amount of observing time, which should increase the probability of 
detecting the types of planets which are known to orbit nearly 20\% of 
these types of stars \citep{johnson10a}.

\subsection{Observations and Data Reduction}

PPPS Doppler measurements are made with the UCLES echelle spectrograph 
\citep{diego:90} at the 3.9-metre Anglo-Australian Telescope (AAT).  
UCLES achieves a resolution of 45,000 with a 1-arcsecond slit.  An 
iodine absorption cell provides wavelength calibration from 5000 to 
6200\,\AA.  The spectrograph point-spread function and wavelength 
calibration are derived from the iodine absorption lines embedded on 
every pixel of the spectrum by the cell \citep{val:95,BuMaWi96}.  The 
result is a precision Doppler velocity estimate for each epoch, along 
with an internal uncertainty estimate, which includes the effects of 
photon-counting uncertainties, residual errors in the spectrograph PSF 
model, and variation in the underlying spectrum between the iodine-free 
template, and epoch spectra observed through the iodine cell.  The 
photon-weighted mid-time of each exposure is determined by an exposure 
meter.  All velocities are measured relative to the zero-point defined 
by the template observation.  Velocities are obtained using the 
\textit{Austral} code as first discussed in \citet{endl00}.  
\textit{Austral} is a proven Doppler code which has been used by the 
McDonald Observatory planet search programs for nearly 10 years 
(e.g.~Endl et al.~2004, 2006; Wittenmyer et al.~2009).

Observations for the PPPS began at the Anglo-Australian Telescope in 
2009 February.  Since its inception, the program has received 20 nights 
per year, of which approximately 50\% have resulted in usable data.  We 
aim for a signal-to-noise (S/N) of 100 at 5500\,\AA\ per spectral pixel 
each epoch, resulting in exposure times ranging from 100\,s up to a 
maximum of 20 minutes.

We have observed 7~CMa on 21 epochs, and an iodine-free template 
observation was obtained on 2010 Jan 30.  Since 7~CMa is an extremely 
bright star, exposure times ranged from 100 to 500 s, with a resulting 
S/N of $\sim$200-300 per pixel each epoch.  The data span a total of 917 
days, and have a mean internal velocity uncertainty of 6.5\,\ms.

\section{Stellar Parameters of 7 CMa}

7 CMa (=HD~47205, HIP~31592) is one of the brightest stars in the PPPS 
survey ($V=3.95$).  In addition, it is accessible from most sites in 
both hemispheres (RA: 06\,36\,41.038, Dec: -19\,15\,21.17), and so it 
has been well-studied.  Table~\ref{stellarparams} summarises the 
physical parameters of this star.  We have used our iodine-free template 
spectrum to derive spectroscopic stellar parameters, using methods 
described fully in \citet{wang11}.  In brief, 7~CMa is an evolved, 
somewhat metal-rich ([Fe/H]$\sim$\,0.2), intermediate-mass star (1.52 
\Msun) with a low level of activity.  \textit{Hipparcos} observations 
indicate that it is photometrically stable, with a median 
\textit{Hipparcos} magnitude of 4.1200$\pm$0.0004 \citep{vl07, 
perryman97}.

\section{Orbit Fitting and Planetary Parameters}

The AAT data show a root-mean-square (RMS) scatter of 29.5\,\ms\ about 
the mean velocity.  Visual investigation of the data after two years of 
observation revealed a clear sinusoidal trend.  Due to the relative 
paucity of data points ($N=21$) compared to typical radial-velocity 
planet detections ($N\gtsimeq40$), the traditional periodogram approach 
does not produce reliable estimates of statistical significance.  
Rather, since the periodic signal is readily apparent by eye, we used a 
genetic algorithm \citep{charbonneau95} to determine Keplerian orbital 
parameters.  Those parameters were then used as initial inputs for a 
standard least-squares fitting routine.  Our previous experience with 
genetic algorithms \citep{cochran07, tinney11} has shown that the 
solution ``evolves'' quite rapidly toward a sharp $\chi^2$ minimum when 
brought to bear on data containing a real and coherent Keplerian signal.  
Indeed, with an allowed period range of 600-800 days, the algorithm 
converged on a solution with a period of 769 days and a small 
eccentricity $e=0.23$.  We then used the \textit{GaussFit} code 
\citep{jefferys87} to obtain a Keplerian model fit for the planet.  For 
the final orbit fitting, we added 5 \ms\ of jitter in quadrature to the 
internal uncertainties of the data shown in Table~\ref{aatvels}.  The 
jitter estimate of 5 \ms\ is derived from \citet{johnson10a}.  In that 
work, 382 velocity measurements from 72 stable stars in the Lick and 
Keck survey of ``retired A stars'' were used to make an empirical jitter 
estimate which was then applied to the seven planet-host stars described 
therein.  It would be ideal to estimate jitter for PPPS stars using the 
same methodology, but at this time, we have insufficient data on 
radial-velocity stable stars to make a statistically meaningful 
estimate.  This is due to the short time baseline (2.5~yr) and limited 
available data on a smaller number of stars in the PPPS as compared to 
the well-established Lick \& Keck survey.  However, we consider the 
5\ms\ jitter estimate to be a reasonable approximation for PPPS targets 
due to the similarity in physical properties to the \citet{johnson10a} 
sample.  We also note that there is substantial ($\gtsimeq$50\%) 
uncertainty in the estimation of radial-velocity jitter 
\citep{wright05}.

\begin{figure}
\plotone{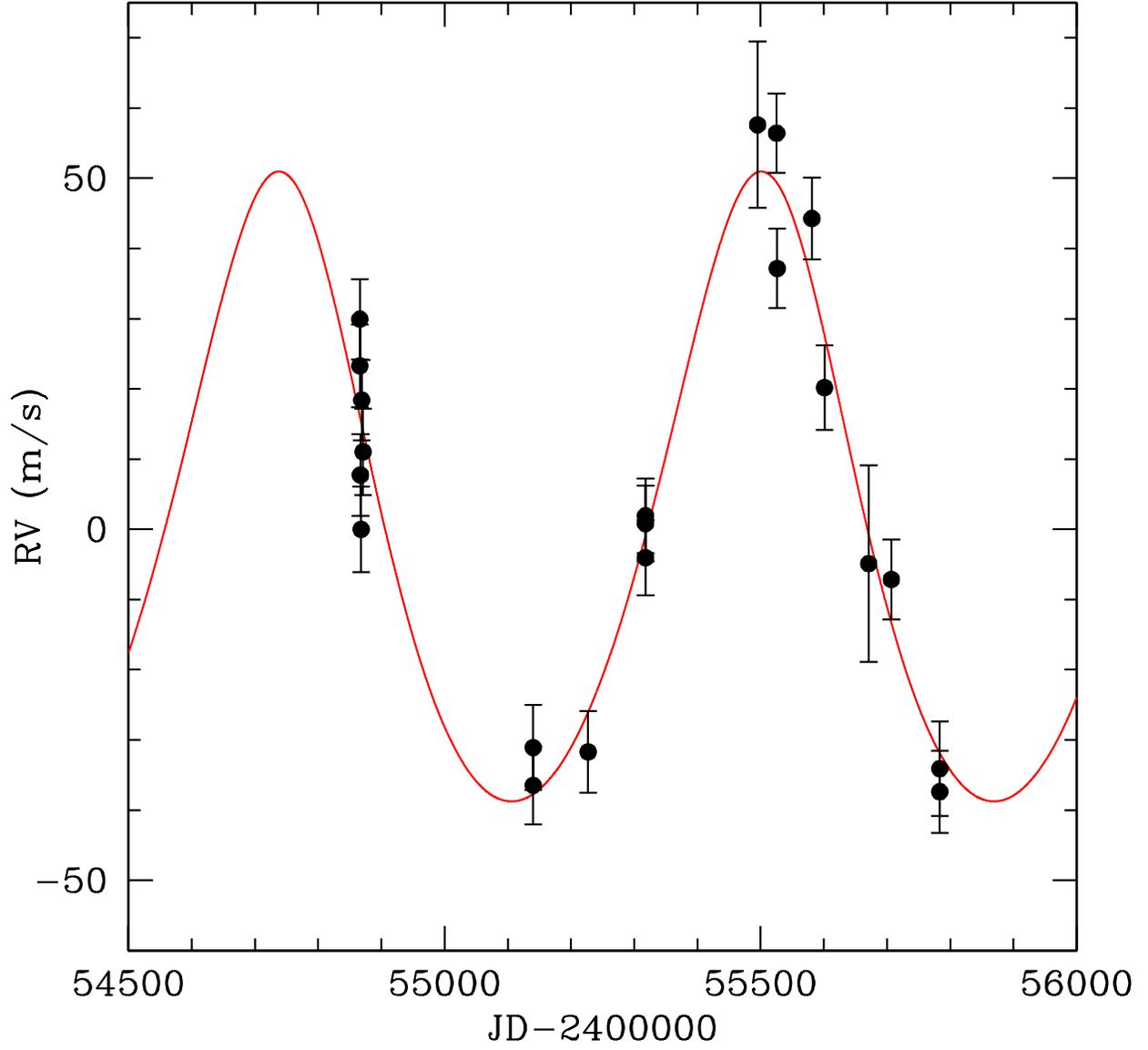}
\caption{Radial-velocity data and Keplerian orbit fit for 7~CMa.  The 
RMS scatter about the fit is 7.5\,\ms, consistent with the mean 
uncertainty of 8.3\,\ms\ (including 5\,\ms\ of jitter added in 
quadrature).}
\label{planetfit}
\end{figure}

\begin{figure}
\epsscale{0.7}
\plotone{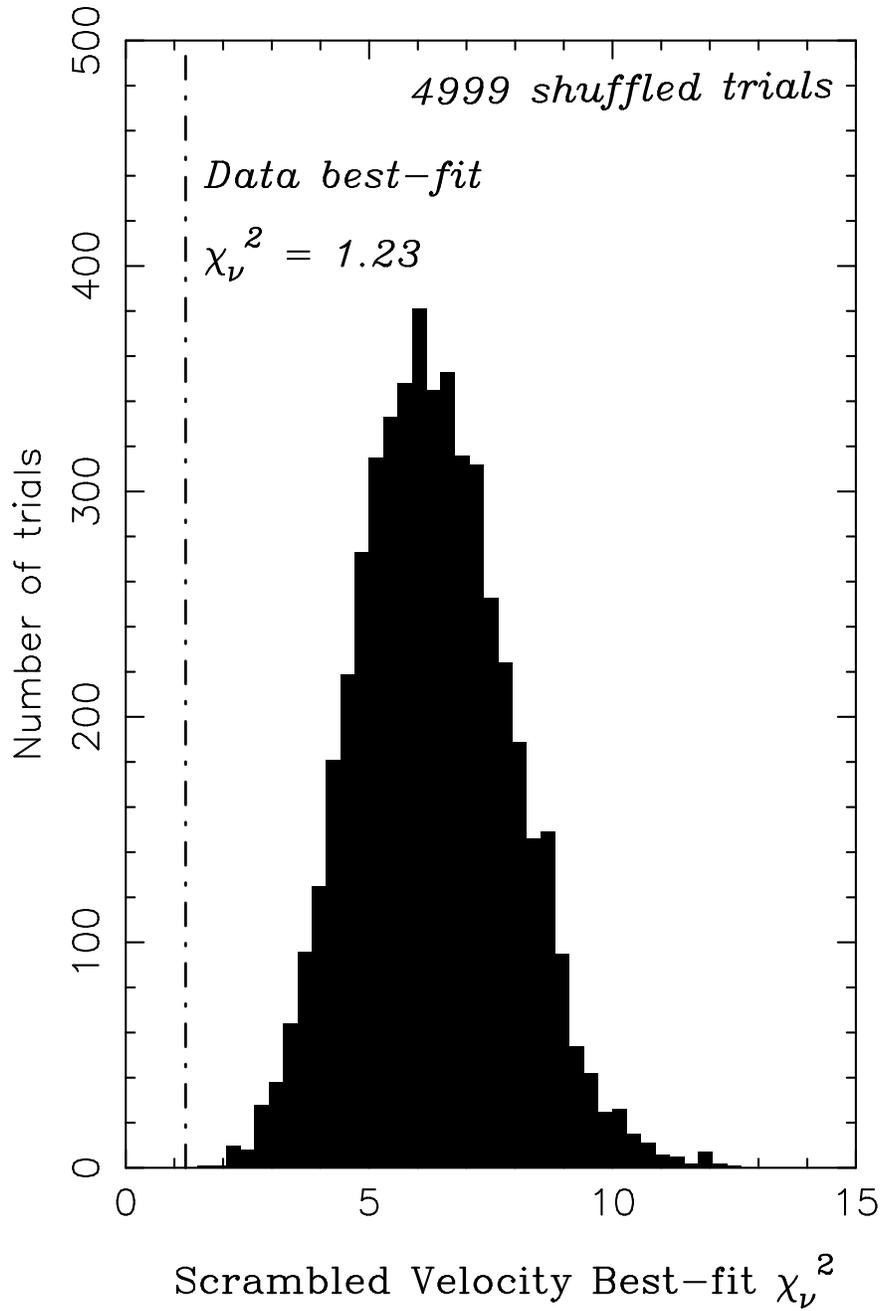}
\caption{Results of 4999 trials in which the velocity data for 7~CMa 
were scrambled amongst the observation epochs.  The reduced $\chi^2$ of 
the original data is shown as a dashed vertical line.  None of the 
scrambled data sets resulted in a better $\chi^2$, indicating a less 
than 0.02\% chance that the observed variations are due to noise. }
\label{shuffle}
\end{figure}

\begin{figure}
\epsscale{1.0}
\plotone{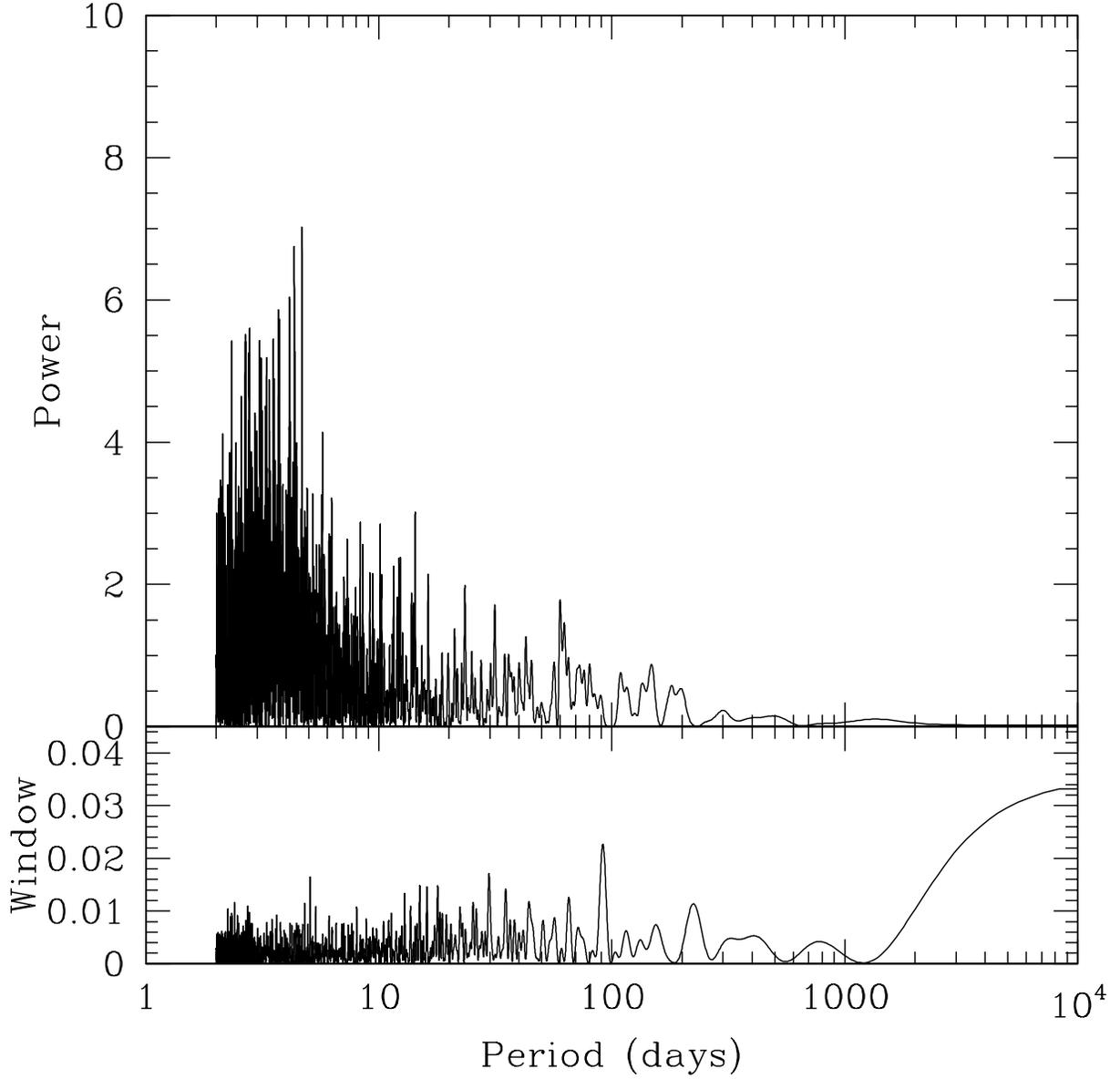}
\caption{Periodogram of the residuals to the Keplerian orbit fit; no 
further signals are evident. }
\label{periodograms}
\end{figure}

Using a stellar mass of 1.52$\pm$0.30 \Msun, we estimate the minimum 
mass m~sin~$i$ to be 2.6$\pm$0.6 \Mjup.  The fit is shown in 
Figure~\ref{planetfit} and the planetary parameters are given in 
Table~\ref{planetparams}.  The residuals of the fit show no evidence for 
additional signals (Figure~\ref{periodograms}).  As a further test of 
the veracity of the planet fit, we used the ``scrambled velocity'' 
approach of \citet{marcy05}.  This technique serves to test the null 
hypothesis that the observed velocity variation is attributable to 
noise.  For this test, we scramble the velocities amongst the 
observation epochs, creating 5000 shuffled data sets.  Then, we perform 
the same least-squares Keplerian orbit-fitting on the shuffled data and 
log the resulting best-fit $\chi^2$.  The results of these trials are 
shown in Figure~\ref{shuffle} -- not one of the scrambled data sets 
achieved a better $\chi^2$ than the planet fit to our original data.  We 
thus conclude that there is a less than 0.02\% probability that the 
detected signal arose by chance from noise.

\section{Discussion}

\subsection{Testing the Planet Hypothesis}

Since many K~giants have intrinsic radial-velocity variations with 
periods of hundreds of days \citep{hekker08}, it is prudent for us to 
further examine the planet hypothesis for 7~CMa to ensure that the 
observed velocity variations are not associated with known activity and 
rotational cycles.  The first and simplest test is to combine the 
available estimates of the star's radius and v~sin~$i$ minimum 
rotational velocity to obtain a maximum rotation period.  Using the 
values for these quantities given in Table~\ref{stellarparams}, this 
yields a maximum $P_{rot}=116$ days (for v~sin~$i$=1.0 \kms, Massarotti 
et al.~2008).  Unfortunately, neither estimate of v~sin~$i$ has an 
uncertainty, but if we apply a typical uncertainty of 1 \kms, then 
maximum rotation periods shorter than 776 days fall within the 1$\sigma$ 
range.

\begin{figure}
\plotone{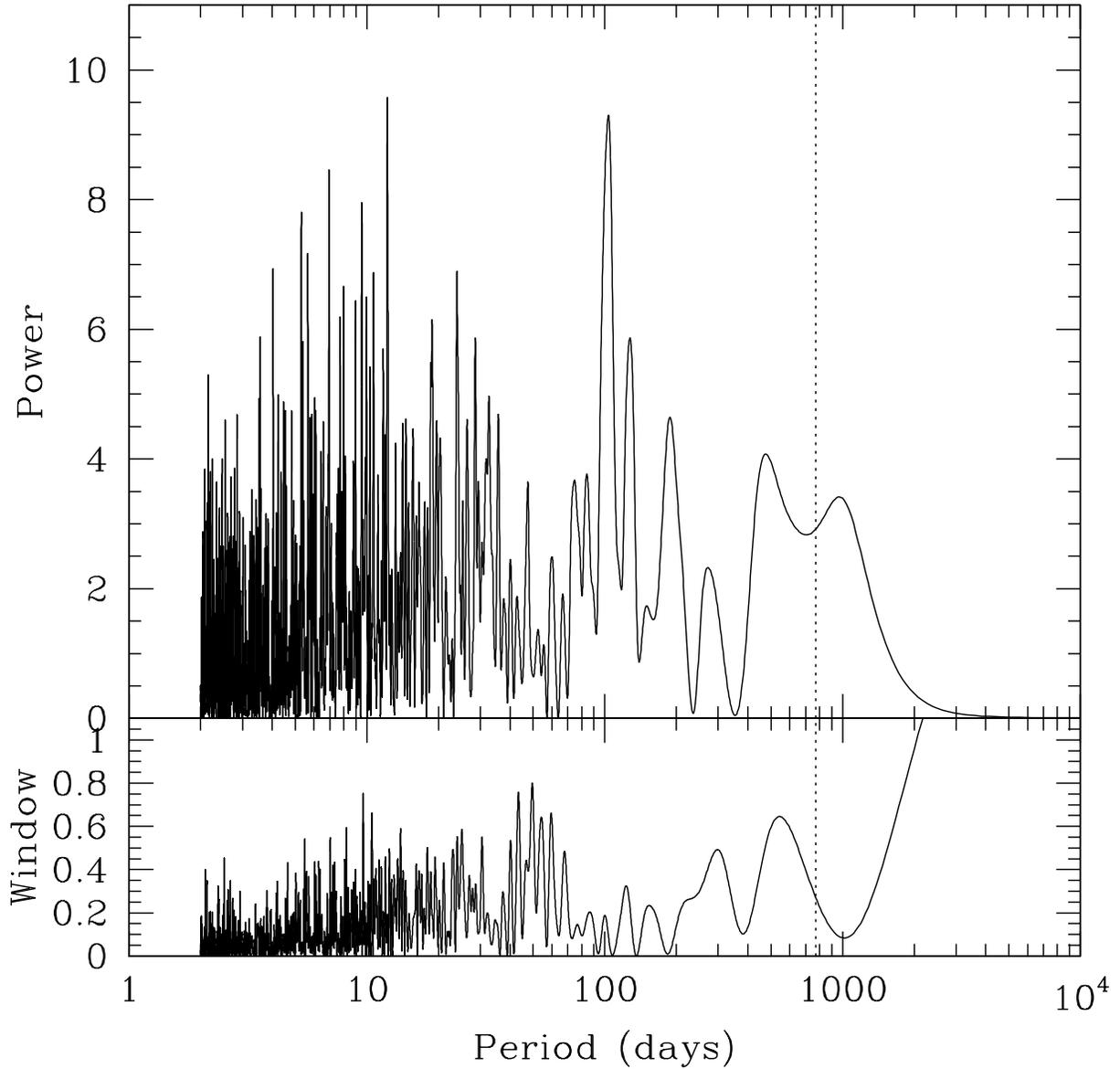}
\caption{Periodogram of \textit{Hipparcos} photometry for 7~CMa 
($N=168$).  The two highest peaks are at periods of 12.2 and 103.7 days.  
The vertical dashed line indicates the 763-day period of the planet; 
there is no significant periodicity in the photometry near this period.}
\label{hipp}
\end{figure}

For a spotted star, the rotation period can be deduced from photometry.  
A periodogram of the \textit{Hipparcos} photometry (after removing one 
outlier which was more than 1 magnitude discrepant) is shown in 
Figure~\ref{hipp}.  Two peaks are evident at periods of 12.2 and 103.7 
days.  We estimate the false-alarm probability using the bootstrap 
randomization method \citep{kurster97}.  The bootstrap method randomly 
shuffles the observations while keeping the times of observation fixed.  
The periodogram of this shuffled data set is then computed and its 
highest peak recorded.  From 10,000 such realizations, we find a 
false-alarm probability of 2.5\% for the peak at 12.2 days, and 3.2\% 
for that at 103.7 days.  At the 763-day period of the candidate planet, 
the bootstrap false-alarm probability is 98.7\%.  The amplitude of the 
photometric variations for either of the two marginally significant 
periods is 20$\pm 6\times 10^{-4}$ mag.  In any case, if these small 
photometric variations are due to the star's rotation, their 
periodicities are clearly well-separated from that of the candidate 
planet.

\begin{figure}
\plottwo{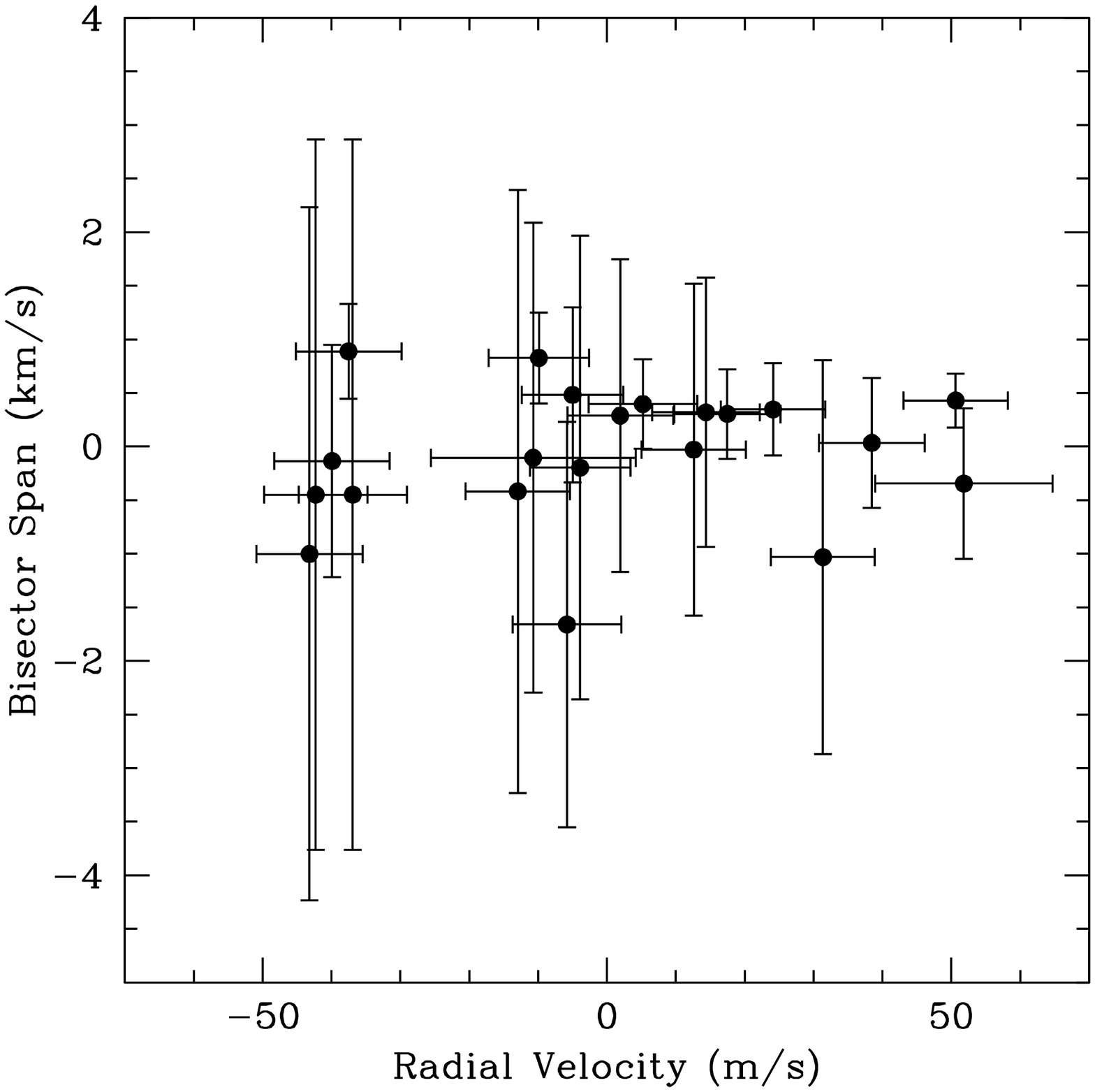}{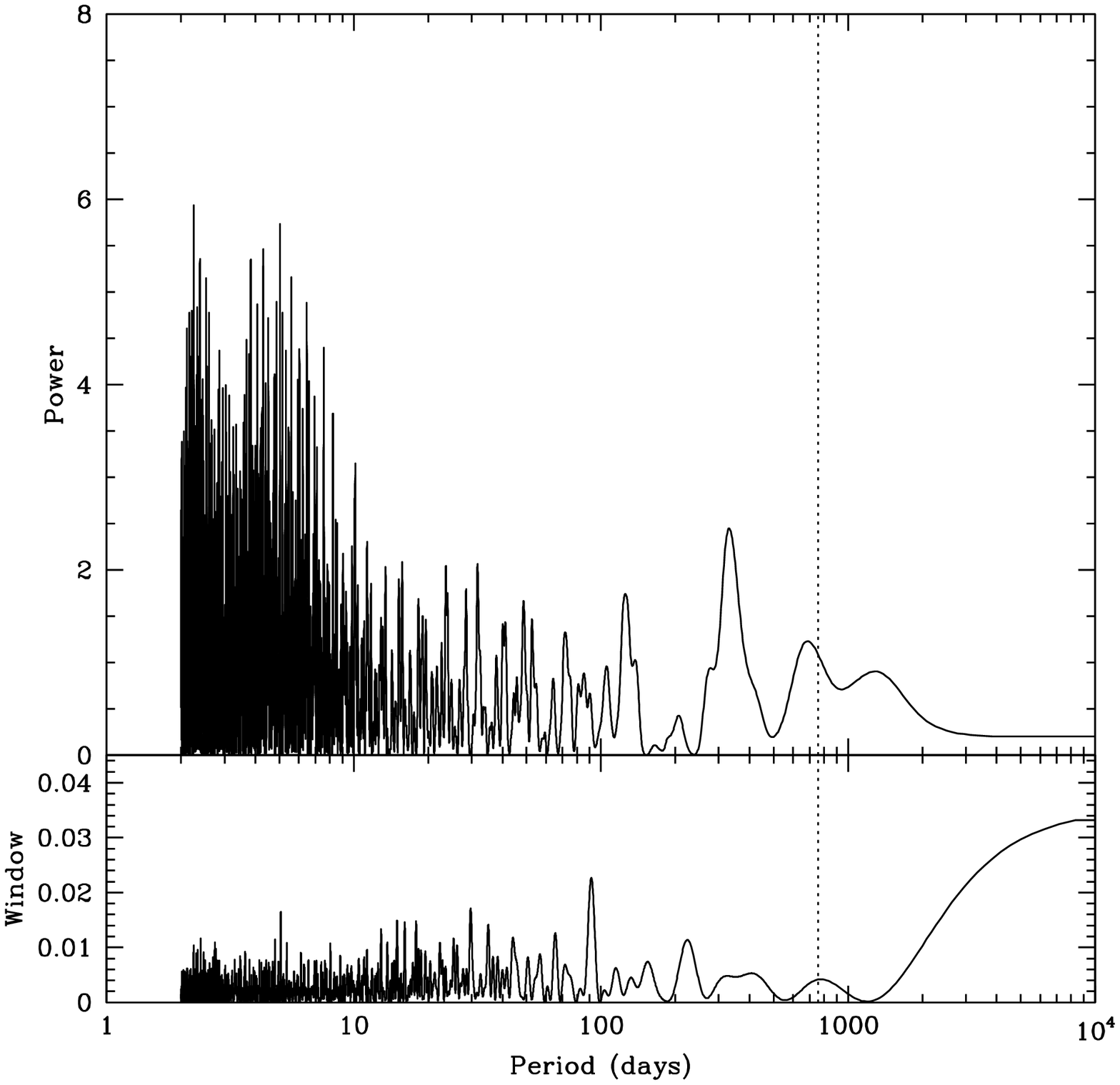}
\caption{Left panel: Bisector velocity span versus radial velocity for 
the 21 observations for 7~CMa.  A correlation would indicate that the 
observed radial-velocity variations were due to an intrinsic stellar 
process rather than an orbiting planet; no correlation is evident.  
Right panel: Periodogram of the bisector velocity spans.  The vertical 
dashed line indicates the 763-day period of the planet; there is no 
significant periodicity near this period. }
\label{bisectors}
\end{figure}

One can argue that the absence of significant variations in the 
\textit{Hipparcos} photometry on the 763-day period is not a complete 
refutation of the starspot hypothesis.  Stars have activity cycles, and 
so there is the possibility that the activity of 7~CMa was at a minimum 
during the \textit{Hipparcos} observations (20 years before the 
radial-velocity observations), but is now at a maximum which, if the 
rotation period were as long as $\sim$763 days, could mimic the signal 
of an orbiting planet.  Line bisector analysis is a fairly common 
technique used by some planet-search programs (e.g.~HARPS) to make sure 
a signal is not due to stellar activity.  Such analysis has the 
advantage of being contemporaneous with the velocity measurements.  We 
note that for the radial-velocity programs using AAT/UCLES, spectra are 
of relatively low resolution ($R=45,000$) and nearly all of the usable 
spectral range is superimposed with iodine lines.  We have computed the 
bisector velocity spans for 8 strong unblended lines redward of the 
iodine region.  The results are shown in Figure~\ref{bisectors}; each 
point represents the mean bisector velocity span of 8 lines, and its 
uncertainty is the standard deviation about the mean value.  While the 
uncertainties are large, it is evident from Figure~\ref{bisectors} that 
the bisector velocity spans are uncorrelated with the radial velocities.  
Furthermore, the right panel of Figure~\ref{bisectors} shows a 
periodogram of the bisector velocity spans, which also indicates no 
periodicity near the 763-day period of the planet.  These independent 
lines of evidence thus lead us to conclude that the radial-velocity 
variations observed in 7~CMa are attributible not to an intrinsic 
stellar process, but to an orbiting giant planet.

\subsection{Conclusions}

\begin{figure}
\plotone{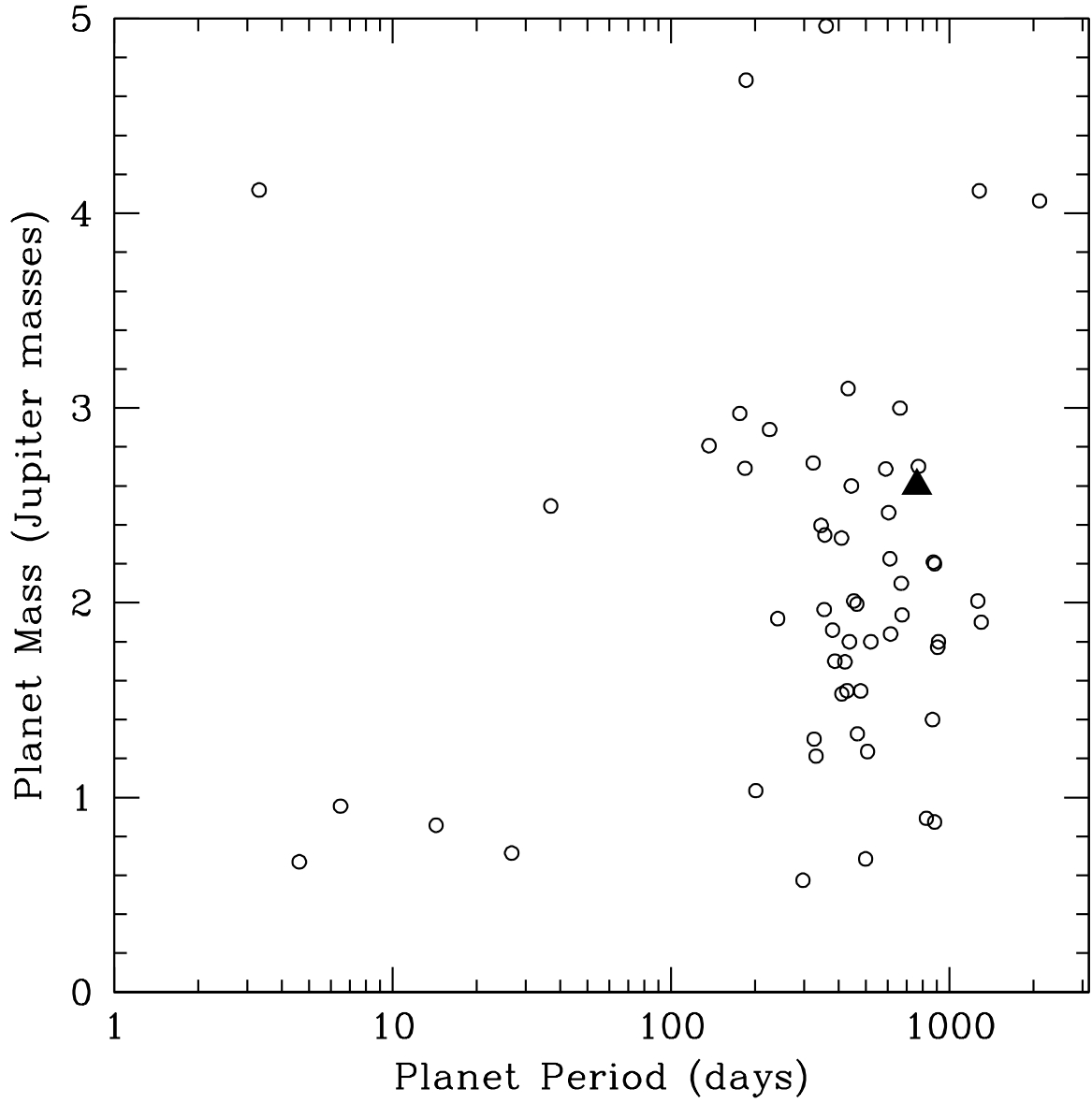}
\caption{Mass-period plot for 83 radial-velocity detected planets 
orbiting stars with $M_{*}>1.3$ \Msun; planet data are from the 
Exoplanet Orbit Database \citep{wright11}.  7~CMa~b is marked as a large 
filled triangle. Its parameters are consistent with other planets 
orbiting intermediate-mass stars. }
\label{compare}
\end{figure}

Using the Anglo-Australian Telescope, we have begun a Southern 
hemisphere search for planets orbiting evolved, intermediate-mass stars.  
Our Pan-Pacific Planet Search team members are based in Australia, 
China, and the US; we now report the first planet detection from our 
ongoing survey.  There is an emerging trend that planets orbiting 
intermediate-mass stars tend to have higher masses and longer periods 
that planets orbiting solar-mass stars \citep{bowler10, johnson10b}.  
With a period of 2.1 years and a minimum mass of 2.6 \Mjup, 7~CMa~b is 
quite similar to other planets known to orbit intermediate-mass stars.  
Figure~\ref{compare} shows the mass and period distribution of all 
radial-velocity detected planets known to orbit stars with 
$M_{*}>1.3$\Msun, with 7~CMa~b plotted as a large filled triangle.  
Given the abundance of planets with $P\gtsimeq$2\,yr known to orbit 
these types of stars, we anticipate that 7~CMa~b is the first of many 
planet detections to come from the PPPS.

\acknowledgements

We gratefully acknowledge the UK and Australian government support of 
the Anglo-Australian Telescope through their PPARC, STFC and DIISR 
funding; STFC grant PP/C000552/1, ARC Grant DP0774000 and travel support 
from the Australian Astronomical Observatory.  RW is grateful to the 
Chinese Academy of Sciences for the support of his stay in Beijing.  RW 
is supported by a UNSW Vice-Chancellor's Fellowship.

We thank the ATAC for the generous allocation of telescope time which 
facilitated this detection.  This research has made use of NASA's 
Astrophysics Data System (ADS), and the SIMBAD database, operated at 
CDS, Strasbourg, France.  This research has made use of the Exoplanet 
Orbit Database and the Exoplanet Data Explorer at exoplanets.org.


\begin{deluxetable}{llll}
\tabletypesize{\scriptsize}
\tablecolumns{4}
\tablewidth{0pt}
\tablecaption{Pan-Pacific Planet Search Target List }
\tablehead{
\colhead{Star} & \colhead{RA} & \colhead{Dec} & \colhead{$V$}\\
 }
\startdata
\label{targetlist}
224910  & 00 01 44.93  & -16 31 54.2  &  7.83   \\
749  & 00 11 38.06  & -49 39 21.2  &  7.91   \\
1817 &  00 22 21.22 &  -50 59 33.4 &   6.68   \\
2643 &  00 29 54.99 &  -32 16 23.8 &   8.15   \\
4145 &  00 43 50.09 &  -12 00 40.7 &   6.01   \\
5676 &  00 58 12.43 &  -25 52 35.8 &   7.89   \\
5877 &  00 59 19.25 &  -58 24 17.2 &   7.78   \\
6037 & 01 01 38.60 & -16 15 55.3 &  6.47  \\
7931 & 01 18 32.20 & -28 43 58.7 &  7.89   \\
9218 & 01 30 13.84 & -28 51 55.5 &  7.96  \\
9925 & 01 35 43.11 & -53 11 59.8 &  7.82  \\
10731 & 01 43 28.25 & -56 14 04.1 &  7.97   \\
11343 & 01 50 06.22 & -54 27 53.5 &  7.88   \\
11653 & 01 53 00.51 & -52 41 30.1 &  7.91   \\
12974 & 02 06 56.07 & -01 49 25.2 &  7.49  \\
13471 & 02 10 54.48 &  -32 03 42.9 &  7.65  \\
13652 & 02 12 34.49 & -26 19 20.7 &  7.92  \\
14805 & 02 20 43.08 & -62 32 45.8 &  7.68    \\
14791 & 02 22 07.00 & -36 06 23.8 &  7.87  \\
15414 & 02 26 12.48 & -62 55 05.2 &  7.92   \\
19810 & 03 10 51.47 & -11 07 29.4 &  7.22   \\
20035 & 03 11 57.63 & -39 21 57.1 &  6.98  \\
20924 & 03 21 58.79 & -15 27 31.4 &  7.26   \\
24316 & 03 51 32.65 & -17 09 58.8 &  7.71  \\
25069 & 03 58 52.42 & -05 28 10.3 &  5.85  \\
28901 & 04 32 06.76 & -28 48 22.0 &  7.42  \\
29399 & 04 33 34.10 & -62 49 25.1 &  5.79  \\
31860 & 04 57 46.42 & -34 53 32.3 &  7.60  \\
34851 & 05 12 02.21 & -75 21 37.6 &  7.85  \\
33844 & 05 12 36.08 & -14 57 04.3 &  7.29  \\
37763 & 05 31 52.66 & -76 20 30.0 &  5.18  \\
39281 & 05 48 34.16 & -53 40 34.1 &  7.85  \\
40409 & 05 54 05.90 & -63 05 27.7 &  4.65  \\
43429 & 06 15 17.71 & -18 28 37.2 &  5.99   \\
46262 & 06 20 53.87 & -79 04 00.5 &  7.31  \\
47141 & 06 36 05.42 & -24 51 57.8 &  7.45   \\
47205 & 06 36 41.00 & -19 15 20.6 &  3.95  \\
51268 & 06 53 33.56 & -54 52 59.3 &  7.97   \\
58540 & 07 22 57.03 & -55 34 38.8 &  6.89   \\
59663 & 07 25 09.11 & -70 24 13.9 &  7.75   \\
67644 & 08 06 20.28 & -54 02 45.9 &  7.97   \\
72467 & 08 32 01.89 & -29 22 15.1 &  7.59   \\
76321 & 08 54 57.73 & -15 46 45.9 &  7.10   \\
76437 & 08 55 01.65 & -34 08 35.0 &  7.15    \\
76920 & 08 55 16.78 & -67 15 55.9 &  7.83   \\
80275 & 09 17 46.62 & -35 41 23.8 &  7.70   \\
81410 & 09 24 49.04 & -23 49 34.4 &  7.35   \\
84070 & 09 41 13.31 & -46 22 55.1 &  7.88   \\
85128 & 09 43 01.74 & -79 35 30.2 &  7.30   \\
85035 & 09 48 47.03 & -19 18 48.6 &  7.02   \\
87089 & 10 00 34.00 & -61 45 31.3 &  7.93   \\
86950 & 10 01 37.61 & -17 19 58.8 &  7.47   \\
HIP50638 & 10 20 33.31 & -23 38 25.4 &  7.54   \\
94386 & 10 53 32.86 & -15 26 44.5 &  6.34   \\
98516 & 11 19 47.64 & -28 11 19.6 &  7.06  \\
98579 & 11 20 19.05 & -28 19 56.1 &  6.68  \\
100939 & 11 36 48.17 & -37 02 20.5 &  7.94   \\
104358 & 12 01 00.72 & -26 28 47.2 &  7.76    \\
104704 & 12 03 22.27 & -55 19 17.0 &  7.49    \\
104819 & 12 04 11.05 & -22 22 15.6 &  7.93    \\
105096 & 12 06 01.36 & -54 15 28.1 &  7.03   \\
108991 & 12 31 38.47 & -30 58 54.8 &  6.73   \\
109866 & 12 38 49.98 & -62 01 54.1 &  7.76   \\
110238 & 12 40 59.85 & -31 44 15.9 &  7.70   \\
114899 & 13 14 26.28 & -54 57 43.6 &  7.99   \\
115066 & 13 15 04.35 & -30 10 53.0 &  7.83   \\
115202 & 13 15 58.58 & -19 56 34.2 &  5.21   \\
121056 & 13 53 52.27 & -35 18 51.1 &  6.17   \\
121156 & 13 54 16.75 & -28 34 09.9 &  6.05   \\
121930 & 13 59 46.00 & -50 13 40.5 &  7.58   \\
124087 & 14 11 45.85 & -19 01 03.2 &  7.74   \\
125774 & 14 21 49.23 & -10 40 00.5 &  7.99   \\
126105 & 14 24 00.33 & -19 48 02.8 &  7.32    \\
130048 & 14 46 13.51 & -07 47 48.9 &  7.14    \\
131182 & 14 52 21.07 & -11 28 22.7 &  7.95   \\
132396 & 15 00 00.35 & -36 01 49.7 &  6.94    \\
133166 & 15 04 36.20 & -43 53 50.4 &  7.92   \\
133670 & 15 06 27.10 & -22 01 54.1 &  6.13    \\
134443 & 15 11 31.92 & -45 16 44.7 &  7.38    \\
134692 & 15 14 59.90 & -66 53 36.5 &  7.91   \\
135760 & 15 18 17.43 & -41 25 13.0 &  7.05   \\
136295 & 15 20 41.44 & -25 59 24.0 &  7.11   \\
136905 & 15 23 26.06 & -06 36 36.7 &  7.29   \\
137115 & 15 24 57.58 & -22 02 37.1 &  7.65   \\
137164 & 15 27 45.75 & -63 01 14.1 &  7.44   \\
136135 & 15 27 47.09 & -79 18 22.4 &  7.61    \\
138061 & 15 29 59.95 & -12 46 35.6 &  7.78   \\
138716 & 15 34 10.52 & -10 03 50.3 &  4.61   \\
138973 & 15 36 08.23 & -21 44 46.6 &  7.72    \\
142132 & 15 54 24.55 & -41 10 22.3 &  7.70    \\
142384 & 15 55 56.46 & -40 47 11.0 &  7.41   \\
143561 & 16 02 45.22 & -42 30 25.5 &  7.97   \\
144073 & 16 05 01.36 & -26 56 52.0 &  7.60    \\
145428 & 16 11 51.34 & -25 53 00.3 &  7.73  \\
148760 & 16 31 22.87 & -26 32 15.2 &  6.07   \\
153438 & 17 00 29.72 & -21 27 41.3 &  7.35   \\
153937 & 17 06 11.92 & -60 25 14.8 &  7.43   \\
154250 & 17 06 40.99 & -48 00 43.5 &  7.96   \\
155233 & 17 11 04.37 & -20 39 15.2 &  6.81   \\
154556 & 17 12 19.85 & -70 43 15.2 &  6.21   \\
159743 & 17 37 01.69 & -18 59 30.9 &  7.45   \\
162030 & 17 49 57.49 & -24 12 25.1 &  7.02   \\
166309 & 18 11 15.84 & -29 38 22.4 &  7.61   \\
166476 & 18 14 20.10 & -58 42 20.9 &  7.81    \\
170707 & 18 33 33.37 & -50 12 41.5 &  7.75   \\
170286 & 18 35 02.98 & -71 19 26.1 &  7.72   \\
173902 & 18 49 17.14 & -34 44 56.0 &  6.59   \\
175905 & 18 57 35.98 & -00 31 34.6 &  7.66   \\
176002 & 19 00 01.36 & -43 20 49.7 &  7.92   \\
175304 & 19 03 29.12 & -76 06 54.8 &  7.75    \\
177897 & 19 08 42.79 & -45 04 34.3 &  7.74   \\
176794 & 19 10 44.78 & -76 24 15.7 &  6.94   \\
181342 & 19 21 04.26 & -23 37 10.2 &  7.55  \\
181809 & 19 22 40.30 & -20 38 33.6 &  6.72   \\
188981 & 19 58 56.37 & -30 32 17.7 &  6.27    \\
191067 & 20 08 01.75 & -00 40 40.9 &  5.97    \\
196676 & 20 39 05.86 & -04 55 46.2 &  6.46   \\
199809 & 21 00 19.00 & -27 20 35.9 &  7.93   \\
200073 & 21 02 27.05 & -38 31 50.0 &  5.93   \\
201931 & 21 14 16.67 & -45 46 57.2 &  6.89    \\
204073 & 21 26 25.07 & -12 05 42.0 &  6.70  \\
204057 & 21 26 27.06 & -15 14 42.9 &  7.97   \\
204203 & 21 27 29.20 & -20 12 45.2 &  7.84    \\
205577 & 21 36 43.65 & -21 30 09.8 &  7.93   \\
205972 & 21 39 15.19 & -13 53 41.0 &  7.25    \\
205478 & 21 41 28.47 & -77 23 22.1 &  3.73   \\
208431 & 21 56 47.43 & -28 49 03.3 &  7.91   \\
208791 & 21 59 01.31 & -11 17 03.4 &  7.79    \\
214573 & 22 40 07.11 & -49 35 53.2 &  7.37  \\
215005 & 22 42 45.92 & -37 20 43.7 &  7.93   \\
216640 & 22 54 45.60 & -16 16 18.3 &  5.53   \\
216643 & 22 55 11.14 & -46 40 43.9 &  7.53    \\
218266 & 23 07 11.97 & -45 50 33.2 &  7.92    \\
219553 & 23 16 49.69 & -21 12 10.7 &  7.25    \\
222076 & 23 38 08.10 & -70 54 12.3 &  7.47   \\
222768 & 23 43 32.71 & -22 54 07.7 &  7.81   \\
223301 & 23 48 28.19 & -11 30 31.4 &  7.60   \\
223860 & 23 53 13.59 & -11 00 52.6 &  7.66   \\
\enddata
\end{deluxetable}

\begin{deluxetable}{lll}
\tabletypesize{\scriptsize}
\tablecolumns{3}
\tablewidth{0pt}
\tablecaption{Stellar Parameters for 7 CMa }
\tablehead{
\colhead{Parameter} & \colhead{Value} & \colhead{Reference}\\
\colhead{} & \colhead{} & \colhead{(\ms)}
 }
\startdata
\label{stellarparams}
Spec.~Type & K1 III & \citet{gray06} \\
$M_V$ & 2.46$\pm$0.03 & \citet{dasilva06} \\
$B-V$ & 1.037$\pm$0.041 & \citet{vl07} \\
Mass (\Msun) & 1.52$\pm$0.30 & This work \\ 
   & 1.32$\pm$0.12  & \citet{dasilva06} \\ 
Radius (\Rsun) & 2.3$\pm$0.1 & This work \\
Luminosity (\Lsun) & 11.3$\pm$0.3 & This work \\
Distance (pc) & 19.75$\pm$0.09 & \citet{vl07} \\
V sin $i$ (\kms) & 1.15 & \citet{hekker07}  \\ 
  & 1.0 & \citet{ma08} \\
$S_{MW}\tablenotemark{a}$ & 0.132 & \citet{gray06} \\
$[Fe/H]$ & 0.21$\pm$0.10 & This work \\ 
  & 0.18$\pm$0.1 & \citet{dasilva06} \\
  & 0.21 & \citet{hekker07} \\
$v_{micro}$ (\kms) & 1.32$\pm$0.10 & This work \\ 
   & 1.30 & \citet{dasilva06} \\
   & 1.45 & \citet{hekker07} \\
   & 1.0  & \citet{gray06} \\
$T_{eff}$ (K) & 4792$\pm$100 & This work \\ 
   & 4744$\pm$70 & \citet{dasilva06} \\
   & 4830 & \citet{hekker07} \\
   & 4799 & \citet{gray06} \\
log $g$ & 3.25$\pm$0.10 & This work \\ 
 & 3.11$\pm$0.07 & \citet{dasilva06} \\
 & 3.40 & \citet{hekker07} \\
 & 3.05 & \citet{gray06} \\
\enddata
\tablenotetext{a}{Mount Wilson S-index}
\end{deluxetable}

\begin{deluxetable}{ll}
\tabletypesize{\scriptsize}
\tablecolumns{2}
\tablewidth{0pt}
\tablecaption{7 CMa Planetary Parameters }
\tablehead{
\colhead{Parameter} & \colhead{Estimate }
 }
\startdata
\label{planetparams}
Period (days) & 763$\pm$17 \\
Eccentricity & 0.14$\pm$0.06 \\
$\omega$ (degrees) & 12$\pm$41  \\
$K$ (\ms) & 44.9$\pm$4.0  \\
$T_0$ (JD-2400000) & 55520$\pm$89  \\
M sin $i$ (\Mjup) & 2.6$\pm$0.6 \\
$a$ (AU) & 1.9$\pm$0.1  \\
\hline
RMS of fit (\ms) & 7.5 \\
$N$ & 21 \\
\enddata
\end{deluxetable}


\begin{deluxetable}{lrr}
\tabletypesize{\scriptsize}
\tablecolumns{3}
\tablewidth{0pt}
\tablecaption{AAT Radial Velocities for 7 CMa}
\tablehead{
\colhead{JD-2400000} & \colhead{Velocity (\ms)} & \colhead{Uncertainty
(\ms)}}
\startdata
\label{aatvels}
54866.09800  &     24.11  &    5.73  \\
54866.10107  &     17.47  &    5.86  \\
54866.94000  &      1.93  &    5.81  \\
54867.91576  &     -5.81  &    6.09  \\
54869.08575  &     12.59  &    5.72  \\
54871.03478  &      5.22  &    6.13  \\
55140.18702  &    -42.28  &    5.56  \\
55140.19229  &    -36.90  &    6.07  \\
55227.06602  &    -37.51  &    5.82  \\
55317.85529  &     -4.99  &    5.40  \\
55317.85839  &     -9.87  &    5.33  \\
55317.86143  &     -3.89  &    5.31  \\
55495.13719  &     51.81  &   11.84  \\
55525.22369  &     50.62  &    5.68  \\
55526.21028  &     31.35  &    5.66  \\
55581.09317  &     38.45  &    5.83  \\
55601.00002  &     14.37  &    6.01  \\
55670.87749  &    -10.70  &   13.99  \\
55706.84304  &    -12.95  &    5.68  \\
55783.30394  &    -43.20  &    5.85  \\
55783.31112  &    -39.91  &    6.72  \\
\enddata
\end{deluxetable}


\end{document}